\documentclass[prl,twocolumn,showkeys,showpacs,10pt,aps,superscriptaddress]{revtex4-2}
\usepackage[english]{babel}
\usepackage{amsmath}
\usepackage{dcolumn}
\usepackage{amssymb}
\usepackage{textcomp}
\usepackage[utf8]{inputenc}
\usepackage{subfig}
\usepackage{hyperref}
\usepackage[justification=raggedright]{caption}
\usepackage{tikz}
\usepackage{nicefrac}
\usepackage{siunitx}
\usepackage{braket}
\usepackage{float}
\usepackage{import}

\newcommand{\out}{{\text{out}}}
\newcommand{\inp}{{\text{in}}}
\newcommand{\Ib}{\bar{I}}

\newcommand{\be}{\begin{equation}}
\newcommand{\ee}{\end{equation}}
\newcommand{\bea}{\begin{eqnarray}}
\newcommand{\eea}{\end{eqnarray}}

\begin{document}
\title{
Silent White Light
}

\author{Kai Niklas Hansmann}
\email{kai.hansmann@physik.tu-darmstadt.de}
\affiliation{Technical University of Darmstadt, Institute of Applied Physics, Hochschulstrasse 4a, Darmstadt D-64289, Germany}
\author{Franziska Dommermuth}
\affiliation{Technical University of Darmstadt, Institute of Applied Physics, Hochschulstrasse 4a, Darmstadt D-64289, Germany}
\author{Wolfgang Els\"a\ss er} 
\affiliation{Technical University of Darmstadt, Institute of Applied Physics, Hochschulstrasse 4a, Darmstadt D-64289, Germany}
\affiliation{
School of Physics, Trinity College Dublin, Dublin 2, Ireland}
\affiliation{Istituto di elettronica e di ingegneria dell'informazione e delle telecomunicazioni (IEIIT) del Consiglio Nazionale delle Richerche (CNR), Politecnico di Torino, Italy}
\author{Reinhold Walser}
\affiliation{Technical University of Darmstadt, Institute of Applied Physics, Hochschulstrasse 4a, Darmstadt D-64289, Germany}

\date{\today}


\begin{abstract}
We investigate the intra-waveguide statistics manipulation of broadband light by combining semiconductor quantum dot physics with quantum optics. By cooling a quantum dot superluminescent diode to liquid nitrogen temperature of $77$\,K, Blazek \emph{et al.} [Phys. Rev. A \textbf{84}, 63840 (2011)] have demonstrated a temperature-dependent reduction of the second-order intensity correlation coefficient from two for thermal amplified spontaneous emission  light to $g^{(2)}(T=\SI{190}{\kelvin})\approx \num{1.33}$. Here, we model the broadband photon statistics assuming amplified spontaneous emission radiation in a pumped, saturable quantum dot gain medium. We demonstrate that, by an intensity increase due to the quantum dot occupation dynamics via the temperature-tuned quasi Fermi levels, together with the saturation nonlinearity, a statistics manipulation from thermal Bose-Einstein statistics towards Poissonian statistics can be realized, thus producing "silent white light". Such intensity-noise reduced broadband radiation is relevant for many applications like optical coherence tomography, optical communication or optical tweezers.
\end{abstract}

\maketitle


Since the first realization of the laser, there has been a perpetual interest in the quantum fluctuations of light, driven both by fundamental and practical interest \cite{Pike2010}. This can be best summarized in the famous saying of Rolf Landauer: “The noise is the signal” \cite{Beenakker2003}. In this spirit, lasers and thermal light sources as “original light sources” have been considered as benchmarks due to their second-order correlation coefficient $g^{(2)}(0)=\langle I^2\rangle/\langle I\rangle^2$ of unity and two respectively \cite{Loudon2009}, determined within a Hanbury Brown and Twiss (HBT) experiment \cite{HanburyBrown1956},  \footnote{It is important to note that the  Hanbury Brown and Twiss experiment  was originally conceived to exploit correlations of fluctuations for the determination of star diameters \cite{HanburyBrown1956}. }. Nowadays, a HBT classification in the $g^{(2)}(0)$ scheme is the characteristics for each new light source and it has become central to quantum optical measurements. In particular, novel concepts have been conceived to create light with $g^{(2)}(0)$ beyond unity and two, emphasizing controlling and manipulating light statistics into regimes beyond, i.e. tailoring $g^{(2)}(0)$ on-demand.

Immediately after the advent of the laser, Martienssen and Spiller and Arecchi \cite{Martienssen1964, Arecchi1965} realized the so-called pseudo-thermal light source in 1966. There, scattering of laser light at a rotating diffuser transformed the Poissonian laser photon statistics into that of thermal light \cite{Gatti2008,Magatti2009} exhibiting Bose-Einstein statistics with $g^{(2)}(0)$ of two. 
This philosophy of exploiting Gaussian and non-Gaussian random walk scattering processes in media led to the achievement of well-controlled states of light \cite{Bertolotti1970,Odonnell1982,Schaefer1972,Jakeman1978,Lodahl2005,Hartmann2015a} in the framework of light with super-Poissonian statistics, i.\,e., bunched or even super-bunched photon counting statistics. Later on, microscopic and mesoscopic scattering concepts for the manipulation of the light statistics have been comprehensively investigated and extended to waveguides \cite{Kindermann2002,Balog2006,Kondakci2016,Mork2020}. 
The concept of manipulating and tailoring light states and exploiting their novel properties beneficially and on-demand in quantum metrology applications has also been investigated by applying nonlinear optical processes onto light \cite{Allevi2015,Allevi2017}. Very recently, it has been shown that disordered systems permit manipulation and tuning of the output statistics via deterministic and coherent control. Monochromatic coherent light traversing a disordered photonic medium evolved into a random field whose statistics has been dictated by the disorder level \cite{Kondakci2016,Kondakci2017}. Deterministic control over the photon-number distribution was demonstrated by interfering two coherent beams within a disordered photonic lattice thus enabling the generation of super-thermal and sub-thermal light \cite{Kondakci2017_2}. 
The generation of squeezed states of light \cite{Walls1979} with $g^{(2)}(0)$ below one and even single photon states by means of non-linear optics 
have been the next revolutionary steps
and opened a huge field of applications in sensing \cite{Andersen2016,Horoshko2019,Polzik1992,Vahlbruch2008,Aasi2013,Foster2019}. 
 
Optoelectronic semiconductor-based emitters are bridging these concepts. The method of "quiet pumping" pioneered by Y.~Yamamoto can be understood as manipulating and creating a particularly interesting emission statistics of a semiconductor laser. By transferring the statistics of the quiet, by Coulomb repulsion regulated sub-Poissonian statistics of the injection current into the sub-Poissonian statistics of the emitted photons, squeezed states of light emerge \cite{Yamamoto1992,Machida1987}. In the current evolution of quantum information technologies, quantum dots (QDs) get embedded in optical integrated chips. Semiconductor QDs have been identified as a promising hardware for implementing the basic building blocks, e.g. stationary and flying qubits in the solid state \cite{Ulrich2007,Oulton2015}.
 
More recently, even so-called hybrid light has been discovered by Blazek et al.  \cite{Blazek2011}. They investigated the emission properties of quantum dot superluminescent diodes (QDSLDs). At room temperature, they are light sources with an ultra-broadband emission spectrum (first-order coherence) and a second-order correlation coefficient $g^{(2)}(0)=2.0$. The authors demonstrated experimentally, however, that the intensity fluctuations can be suppressed down to $g^{(2)}(0)=1.33$ while tuning the temperature to $190\,$K. Such emission, being first-order incoherent (spectrally broad-band) and second-order coherent (towards that of a laser), might be called "silent white light" and is of particular interest for applications such as optical coherence tomography \cite{Huang1991}, fiber optic gyroscopy \cite{Bohm1981} and ghost imaging \cite{Pittman1995,Gatti2006}. Previous theoretical investigations of this behavior focused on the quantum nature of the diode material \cite{phdfriedrich2019,Friedrich2020,Hansmann2021}.


Here, we will present a quantitative model for the explanation of these observations \cite{Blazek2011}. It accounts for the self-consistent modification of the photon statistics caused by the nonlinear response of the QD gain medium \cite{Walser1994}, as well as thermally induced occupation of its energy levels \cite{Huang2001,Zhukov1997}. 
This insight will promote further developments, thus paving the avenue for novel, compact and fully integrated light sources.

\begin{figure}[h!]
	\centering
	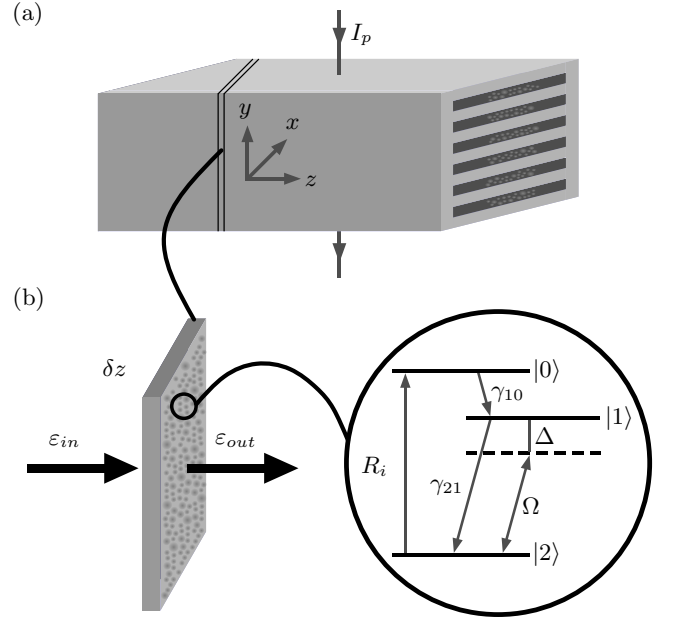
	\caption{
		(a) QDSLD pumped by an electric injection current $I_p$. 
		Layers of wave guides with non-reflecting tilted end-facets and embedded QDs serve as a lossy, nonlinear gain material. 
		(b) Stochastic electric field $\varepsilon(t)$ propagating for a length $\delta z$ through a transversal sheet of inverted three-level QDs with
		decay rates $\gamma_{10}, \gamma_{21}$, an internal pumping rate $R_i$, 
		a Rabi frequency $\Omega$ and a detuning $\Delta$.
		\label{fig:diode_system}}
\end{figure}

Radiation is generated in an active QD gain medium by amplified spontaneous emission (ASE). In contrast to a laser, the wave-guide medium is terminated with non-reflecting tilted end-facets, shown in Fig.~\ref{fig:diode_system} (a). Therefore, there is no feedback mechanism to create coherence. The equilibrium state of the radiation inside the diode is a balance between the nonlinear gain from the QD ensemble, its saturation behavior, the absorption from the passive medium and the emission output coupling of the diode. In the following, these processes are considered in detail for a thin transversal layer of QDs interacting with the radiation field.

According to the Maxwell-Bloch equations \cite{Cohen-Tannoudji1998,Meystre2007}, the slowly varying electric field amplitude $\varepsilon(t)$ passing through a thin sheet $\delta z$ of polarizable matter (see Fig.~\ref{fig:diode_system}(b)) reads
    \begin{align}
        \varepsilon_\out(t)=\eta\varepsilon_\inp(t)+\frac{ik\delta z}{2\varepsilon_0}\mathcal{P}^{(+)}(t).
        \label{eq:inOutField}
    \end{align}
Here, $0<\eta<1$ accounts for scattering losses, $k$ is the carrier wave number of the electric field and $\mathcal{P}^{(+)}$ is the positive frequency part of the polarization. QDs are the active agents embedded in the passive waveguide layers inside the diode. They are modeled as pumped three-level systems (see inset in Fig. \ref{fig:diode_system}(b)). 

The electric field drives the transition between levels $\vert 1\rangle$ and $\vert 2\rangle$ with Rabi frequency $\Omega(t)=d_{21}\varepsilon(t)/\hbar$, where $d_{21}$ is the dipole matrix element. The positive frequency part of the polarization $\mathcal{P}^{(+)}=n d_{12}\rho_{21}$ scales with the density of QDs $n$. In the rate equation limit \cite{Cohen-Tannoudji1998}, coherences $\rho_{ij}$ decay much faster than populations $\rho_{ii}$, which  evolve as
    \begin{align}
        \dot{\rho}_{00}=&-(R_i+\gamma_{10})\rho_{00}+R_i\rho_{22}, \nonumber\\
        \dot{\rho}_{11}=&\gamma_{10}\rho_{00}-(\gamma_{21}+\zeta)\rho_{11}+\zeta\rho_{22}, \label{eq:rateEquations}\\
        \dot{\rho}_{22}=&R_i\rho_{00}+(\gamma_{21}+\zeta)\rho_{11}-(R_i+\zeta)\rho_{22} \nonumber,
    \end{align}
where $\zeta=\vert\Omega\vert^2\mathcal{L}/\gamma$, $\gamma=\gamma_{21}+R_i$ and $\mathcal{L}=(\gamma/2)^2/(\Delta^2+(\gamma/2)^2)$. In this limit, the coherence $\rho_{21}=-(i/2)\Omega\mathcal{D} w$ is instantaneously related to the inversion $w=\rho_{11}-\rho_{22}$ and the stochastic field $\varepsilon(t)$, which is modulated by the lineshape $\mathcal{D}=1/(i\Delta+\gamma/2)$. Hence,
    \begin{align}
        \rho_{21}=-\frac{i}{2}\Omega\frac{w_0 \mathcal{D}}{1+s},\label{eq:rho21stat}
  \end{align}
with the unsaturated inversion $w_0=(R_i(\gamma_{10}-\gamma_{21})-\gamma_{10}\gamma_{21})/(\gamma\gamma_{10}+2R_i\gamma_{21})$ and denoting the saturation parameter $s=\mathcal{L}I/I_s$ as the ratio of intensity $I=\vert\varepsilon\vert^2$ and the saturation intensity $I_s=\hbar^2\gamma(\gamma\gamma_{10}+2R_i\gamma_{21})/\vert d_{21}\vert^2(3R_i+2\gamma_{10})$.

Now, the input-output-relation from Equ. \eqref{eq:inOutField} reads $\varepsilon_\out=(\eta+\alpha)\varepsilon_\inp$ and depends on a nonlinear absorption coefficient $\alpha=\kappa\gamma\mathcal{D}/(1+s)$, the nonlinear saturation parameter $\kappa=\alpha_0k\delta z$ and the linear, resonant absorption coefficient $\alpha_0=n\vert d_{21}\vert^2 w_0/4\hbar\varepsilon_0\gamma$ \cite{Meystre2007}. In the limit of a thin sheet of length $\delta z$, all terms exceeding linear order in $\kappa$ can be neglected and the input-output relation for the intensities is obtained
    \begin{align}
        I_{\out}(I_{\inp})=\left(\eta^2+\frac{4\eta\kappa}{1+s}
        \right)I_{\inp}+
        \mathcal{O}[\delta z^2]. 
        \label{eq:in_out_intensity}
    \end{align}

 As a consequence of the broad \si{THz}-bandwidth of the first-order incoherent light, we can choose the detuning $\Delta=0$. Thus, only three parameters $\eta$, $\kappa$ and $I_s$ remain unspecified in Equ. \eqref{eq:in_out_intensity}.  

Here, we start with a incoherent Gaussian photon statistics as the starting point for the statistics manipulation by the QD saturable medium and the QD level scheme. This implies an exponential probability density for the input intensity $p(I_\inp)$ \cite{Hansmann2021} 
    \begin{align}
	    p(I_\inp)&=e^{-I_\inp/\Ib}/\Ib, &  \langle I_\inp \rangle &=\Ib,
    \end{align}
with an average intensity $\Ib$. Thus, the $n$-th order moments of the output photon intensity are given by 
    \begin{align}
        \langle I_{\out}^n(I_{in})\rangle =\int_0^\infty\text{d} I_\inp
        \;I_{\out}^n(I_\inp)\, p(I_\inp).
    \end{align}

In equilibrium, gain is compensated by loss and defines a self-consistent relation for the intensity \eqref{eq:in_out_intensity}
    \begin{align}
       \langle I_{\out}\rangle= 
       \langle I_{\inp}\rangle=
       \Ib(\eta).
    \end{align}
From this condition, we can  determine the inaccessible loss rate $\eta(\Ib)$ in favor of the equilibrium intensity $\Ib$.
 
\begin{figure}[h]
	\centering
	\scalebox{1.0}{\import{Pictures/}{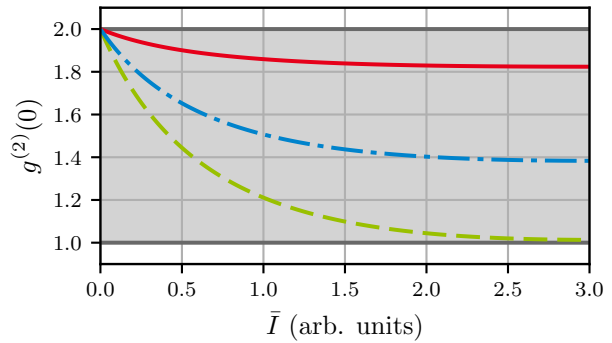}}
	\caption{Central second-order coherence correlation coefficient $g^{(2)}(0)$ as a function of mean intensity $\Ib$, for a saturation intensity $I_s=5$ and for three nonlinear saturation parameters $\kappa$= $0.1$ (red, solid), $0.35$ (blue, dash-dotted), $0.56$ (green, dashed). 
	\label{fig:g2_of_i0}}
\end{figure}

Now, we are able to evaluate the stationary, zero delay time ($\tau=0$) relative intensity-noise correlation function  
    \begin{align}
        g^{(2)}(\tau=0,\Ib)=\lim\limits_{t\rightarrow\infty}\langle I_{\out}(t)^2\rangle/\langle I_\out \rangle^2
    \end{align}
as a measure for the intensity fluctuations. Serendipitously, one can evaluate this expression analytically in terms of 
$u(\mathcal{I})=e^{\mathcal{I}}\Gamma(0,\mathcal{I})$, the incomplete Gamma function $\Gamma(0,\mathcal{I})=\int_{\mathcal{I}}^\infty\text{d}t\;e^{-t}/t$  \cite{NIST:DLMF} and the relative intensity $\mathcal{I}=I_s/\Ib$. Within the thin sheet limit $\mathcal{O}[\delta z^2]$, one finds in good approximation
    \begin{align}
         g^{(2)}(0)=2-8\kappa\mathcal{I}
        \left[1+\mathcal{I}\left(1-2 u(\mathcal{I})\right)-\mathcal{I}^2 u(\mathcal{I})\right]. \label{eq:g2_i0}
    \end{align}
In Fig.~\ref{fig:g2_of_i0}, we depict this intensity correlation $g^{(2)}(0)$ versus the internal QDSLD intensity $\Ib$ for a chosen saturation intensity $I_s$ and for various saturation parameters $\kappa=0.1, 0.35,$ and $0.56$, respectively. In general, $g^{(2)}(0)$ shows a strong decrease with increasing $\Ib$, which is stronger for higher $\kappa$. However, we note that this is only the case if the medium is inverted, i.\,e. $\kappa>0$.

\begin{figure} [b]
\centering
\scalebox{0.9}{
\begingroup%
  \makeatletter%
  \providecommand\color[2][]{%
    \errmessage{(Inkscape) Color is used for the text in Inkscape, but the package 'color.sty' is not loaded}%
    \renewcommand\color[2][]{}%
  }%
  \providecommand\transparent[1]{%
    \errmessage{(Inkscape) Transparency is used (non-zero) for the text in Inkscape, but the package 'transparent.sty' is not loaded}%
    \renewcommand\transparent[1]{}%
  }%
  \providecommand\rotatebox[2]{#2}%
  \newcommand*\fsize{\dimexpr\f@size pt\relax}%
  \newcommand*\lineheight[1]{\fontsize{\fsize}{#1\fsize}\selectfont}%
  \ifx\svgwidth\undefined%
    \setlength{\unitlength}{246bp}%
    \ifx\svgscale\undefined%
      \relax%
    \else%
      \setlength{\unitlength}{\unitlength * \real{\svgscale}}%
    \fi%
  \else%
    \setlength{\unitlength}{\svgwidth}%
  \fi%
  \global\let\svgwidth\undefined%
  \global\let\svgscale\undefined%
  \makeatother%
  \begin{picture}(1,0.61788618)%
    \lineheight{1}%
    \setlength\tabcolsep{0pt}%
    \put(0,0){\includegraphics[width=\unitlength,page=1]{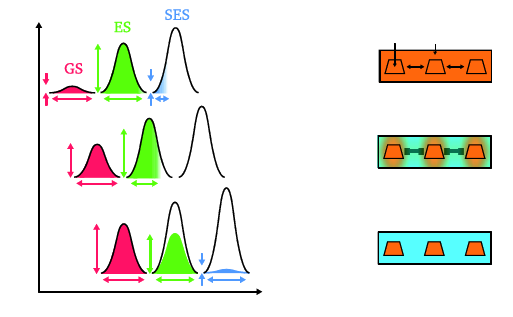}}%
    \put(0.04601392,0.35823017){\color[rgb]{0,0,0}\rotatebox{90}{\makebox(0,0)[lt]{\lineheight{1.25}\smash{\begin{tabular}[t]{l}Population\end{tabular}}}}}%
    \put(0.37171747,0.01084188){\color[rgb]{0,0,0}\makebox(0,0)[lt]{\lineheight{1.25}\smash{\begin{tabular}[t]{l}Energy\end{tabular}}}}%
    \put(0.61593802,0.5124015){\color[rgb]{0,0,0}\makebox(0,0)[t]{\lineheight{1.25}\smash{\begin{tabular}[t]{c}290K\\Fermi\\distribution\end{tabular}}}}%
    \put(0.61593802,0.15455089){\color[rgb]{0,0,0}\makebox(0,0)[t]{\lineheight{1.25}\smash{\begin{tabular}[t]{c}90K\\Random\\distribution\end{tabular}}}}%
    \put(0.61593802,0.34243199){\color[rgb]{0,0,0}\makebox(0,0)[t]{\lineheight{1.25}\smash{\begin{tabular}[t]{c}190K\\Fermi\\distribution\end{tabular}}}}%
    \put(0.77047364,0.55069263){\color[rgb]{0,0,0}\makebox(0,0)[t]{\lineheight{1.25}\smash{\begin{tabular}[t]{c}QD\end{tabular}}}}%
    \put(0.84862338,0.54928137){\color[rgb]{0,0,0}\makebox(0,0)[t]{\lineheight{1.25}\smash{\begin{tabular}[t]{c}WL\end{tabular}}}}%
  \end{picture}%
\endgroup%
}
\caption{Schematic depiction of the occupation of the QD levels as a function of temperature T for the explanation of the experimental measurement of peak power emitted by the diode from \cite{Blazek2011}.}
\label{fig:qd_occupation}
\end{figure}
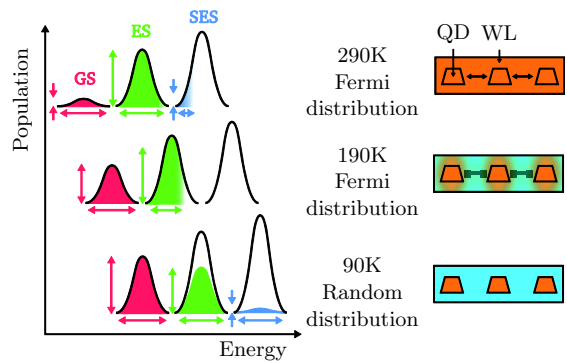

Thermal effects within the QD gain medium influence the carrier population and thus determine the key parameter of $g^{(2)}(0)$ via the generated photon density or the intensity $\Ib$. Its influence on the statistics via the emitted intensity are schematically summarized in Fig.~\ref{fig:qd_occupation}. For the description of this temperature-dependent reduced second-order correlation coefficient, we consult a rate equation model that has been developed previously to describe the threshold currents’ temperature dependence in strongly inhomogeneously broadened QD lasers, reflecting its radiative recombination processes \cite{Huang2001,Zhukov1997}. Thereby, we combine the two worlds of quantum optics and semiconductor quantum dots.

The charge-carrier distribution in semiconductor QD materials depends on temperature, and the mean carrier occupation number for each energy level is obtained by averaging over the whole inhomogeneous dot ensemble. The ingredients of the model are the two confined QD levels, namely the GS (ground state) and the ES (excited state), and the so-called wetting layer, which provides the joint interaction medium for all QDs. Their appropriate interaction is accounted for by relaxation rates, carrier escape processes via thermally activated escape, tunneling, and Auger processes. Finally, all states interact with a bosonic phonon bath. The outcome is the carrier distribution or the population densities entering directly into the radiative photon emission rates.

At room temperature, high-energy phonons induce a global thermal equilibrium of the whole QD ensemble through interaction with the surrounding wetting layer (see Fig. \ref{fig:qd_occupation}). This thermally excites some of the carriers into higher energetic states, leaving some of the lower states unoccupied. Accordingly, the occupation is described by an equilibrium Fermi-Dirac distribution with a global Fermi level for all electron levels.

When the temperature is reduced, the carriers are still uniformly distributed among the individual dots. However, thermal excitations freeze out, the nonradiative losses decrease, and charge-carrier condensation into the globally lowest energy state occurs. This maximizes the occupation numbers of the GS and ES transitions.

At even lower temperatures, this common occupation statistics or global equilibrium collapses. The exchange of carriers between the individual dots breaks down and inside each dot all the energetically lowest states have the same population, irrespective of their energy. This characterizes a so-called random population. The resulting distribution is a non-equilibrium distribution with a “virtual” excitation spectrum obtained by averaging over the whole ensemble, thus reflecting more the energetically inhomogeneous dot distribution. This leads again to a decrease in radiative recombination accompanied by a small increase in linewidth.

In essence, at around $190\,$K, a maximum in the radiative recombination occurs due to the occupation condensation into the globally lowest-lying state that is still described by a Fermi-Dirac distribution. This redistribution of carriers modifies the optical gain properties of the QDSLD that we investigate through temperature-resolved spectral analysis. The spectral peak power extracted from the maximum value of the optical spectra represents an easily accessible indicator for the spectral gain.

\begin{figure}[b]
	\centering
	\scalebox{1.0}{\import{Pictures/}{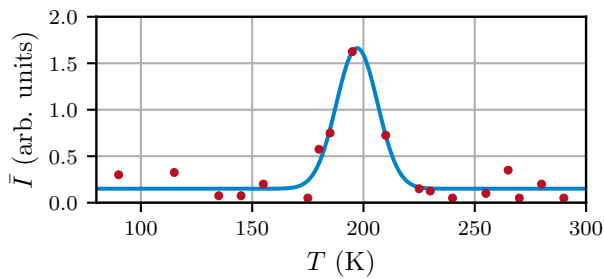}}
	\caption{Mean output intensity $\Ib$ versus temperature $T$. Experimental data (red, \cite{Blazek2011a}) and model (blue, (\ref{eq:i0_gaussian})) yield a maximum intensity at around $190\,$K, implying an increase in diode efficiency at this temperature. \label{fig:i0_of_T}}
\end{figure}

The relative development of the peak power is shown in Fig. \ref{fig:i0_of_T}. In the weakly coupled thermal regime at $190\,$K, we find an increase in peak power compared to room temperature due to the condensation of charge carriers. The local maximum in peak power indicates a larger amplification, which in turn affects the photon emission process. At room temperature, the QDSLD emits amplified spontaneous emission in a delicate balance, where spontaneously emitted photons are amplified moderately. At $190\,$K, the maximum in the spectral gain increases the probability of stimulated emission such that the initial spontaneous emission experiences a stronger amplification. These quasi-stimulated processes reduce the second-order intensity correlation coefficient $g^{(2)}(0)$ and suppress intensity fluctuations \cite{Blazek2007,Shin2010}, thus realizing the exciting hybrid coherent photon states.


The consequence of this behavior is a hierarchy in the contributing QD levels with a peak behavior of the emitted intensity as a function of temperature as illustrated by Fig. \ref{fig:i0_of_T}, which shows the experimental findings of the emitted intensity of the diode as a function of temperature. We can phenomenologically model the emitted power as a temperature-dependent Gaussian function with an offset $\delta I$

    \begin{align}
        \Ib(T)=\Ib e^{-(T-T_0)^2/\sigma^2}+\delta I. \label{eq:i0_gaussian}
    \end{align}
 The experimental data can be fitted well for $\Ib=1.51\pm0.13$, $T_0=(197.1\pm0.9)\,$K, $\sigma=(13.1\pm1.0)\,$K and  $\delta I=0.15\pm0.03$.  


\begin{figure}[b]
\centering
	\scalebox{1.0}{\import{Pictures/}{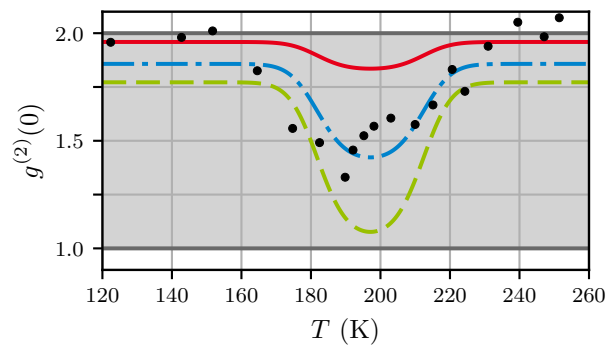}}
	\caption{Central degree of second-order coherence $g^{(2)}(0,\Ib(T))$ versus temperature $T$ for $I_s=5$ and varying saturation parameter values $\kappa$. For $\kappa=0.35$ (blue, dash-dotted), we are able to match the experimental data \cite{Blazek2011}. This agreement deteriorates for $\kappa=0.1$ (red, solid). Within the limits of the model, the intensity noise suppression could  even reach $g^{(2)}(0)=1.07$ for $\kappa=0.56$ (green, dashed).}
	\label{fig:g2_of_T}
\end{figure}

These thermal fitting parameters can be used to construct the temperature-dependent behaviour of $g^{(2)}(0,\Ib(T))$ which is plotted in Fig.~\ref{fig:g2_of_T}. As can be seen in Fig.~\ref{fig:g2_of_T}, all nonlinear saturation parameter combinations $\kappa$ show a suppression of intensity fluctuations around $190\,$K. With the parameters set to $I_s=5$ and $\kappa=0.35$ (blue), we achieve good agreement with the experimental data \cite{Blazek2011}. The calculations using the fitted data from Fig.~\ref{fig:i0_of_T} do not reach a plateau of $g^{(2)}(0)=2.0$ for high and low temperatures. This is due to the finite offset $\delta I=0.15\pm0.03$ of $\Ib(T)$. With the saturation parameters set to $I_s=5$ and $\kappa=0.56$ (green), we are able to produce an intensity noise suppression below the experimentally reported value of $g^{(2)}(0)=1.33$ with a minimum of about $g^{(2)}(0)=1.07$.

 

Having developed a good description of the experimentally observed $g^{(2)}(0)$ reduction of hybrid light, we are now able to search even towards more reduction reaching eventually the Poissonian correlation limit of $g^{(2)} = 1$, still keeping the spectral broadband character. Adjusting our model parameters, we are able to show reductions of $g^{(2)}(0)$ nearly down to a value of $1.07$ for a $\kappa = 0.56$, very close to "real" Poissonian statistics, but now not for a laser but still for a broadband hybrid ASE light source. However, we admit that it is experimentally and technologically quite challenging to find appropriate QD level systems and QDSLD designs preventing stimulated modal emission, thus maintaining low first-order coherence, and avoiding a collapse of the spectral linewidth \cite{Hartmann2013}. 


In conclusion, we have developed a quantum optical model for a thermally-tuned photon statistics transformation of broad-band \si{THz}-wide ASE radiation emitted from a quantum dot superluminescent diode from the thermal Bose-Einstein statistics towards Poissonian statistics, thus producing "silent white light". The two ingredients, nonlinear gain saturation and an increased recombination determined by the temperature dependence of the hierarchy of the quantum dot occupation allowed to account for the experimentally observed findings considering real world parameters. 
These results and their insight will promote further developments thus paving the avenue for novel, compact and fully integrated light sources emitting new tailored quantum states of light with on-demand tailored fluctuation properties opening a huge field of applications in sensing. 


The authors have no conflict of interest to disclose. 
The data that support the findings of this study are available from the corresponding author upon reasonable request.
We gratefully acknowledge supporting discussions with Dr. Martin Blazek. RW and KH are supported by the DLR German Aerospace Center with funds provided by the Federal Ministry for Economic Affairs and Energy (BMWi) under Grant No. No. 50WM2250E.
\bibliography{References.bib}
\end{document}